# High-Resolution Structure of Viruses from Random Snapshots

A. Hosseinizadeh[1*], P. Schwander[1*], A. Dashti[2], R. Fung[1], R.M. D'Souza[2], A. Ourmazd[1]

[1]Dept. of Physics, University of Wisconsin Milwaukee, 1900 E. Kenwood Blvd, Milwaukee, WI 53211

[2]Dept. of Mechanical Engineering, University of Wisconsin Milwaukee, 1900 E. Kenwood Blvd, Milwaukee, WI 53211

* These authors contributed equally

*Corresponding author:* A. Ourmazd; Tel: (414) 229-2610; Email: Ourmazd@uwm.edu

**Abstract**

The advent of the X-ray Free Electron Laser (XFEL) has made it possible to record snapshots of biological entities injected into the X-ray beam before the onset of radiation damage. Algorithmic means must then be used to determine the snapshot orientations and reconstruct the three-dimensional structure of the object. Existing approaches are limited in reconstruction resolution to at best 1/30$^{th}$ of the object diameter, with the computational expense increasing as the eighth power of the ratio of diameter to resolution. We present an approach capable of exploiting object symmetries to recover three-dimensional structure to 1/100$^{th}$ of the object diameter, and thus reconstruct the structure of the satellite tobacco necrosis virus to atomic resolution. Combined with the previously demonstrated capability to operate at ultralow signal, our approach offers the highest reconstruction resolution for XFEL snapshots to date, and provides a potentially powerful alternative route for analysis of data from crystalline and nanocrystalline objects.

*(151 words)*

## Introduction

The advent of X-ray Free Electron Lasers (XFEL's) has opened the unprecedented possibility to record snapshots of biological objects before the onset of radiation damage[1,2]. This recently resulted in de novo determination of protein structure from nanocrystals fabricated in vivo[3]. The ultimate goal, however, remains the determination of the three-dimensional structure of *individual* proteins and viruses[4], and their conformations[5]. This requires the ability to recover structure from an ensemble of ultralow-signal diffraction snapshots of unknown orientation.

This algorithmic challenge was first solved by Bayesian approaches[6,7]. The computational expense of these methods, however, limits their resolution to ~1/10th of the object diameter[6-8], far below the level required for biological impact. Computationally efficient non-Bayesian methods have extended the achievable resolution to 1/30th of the object diameter[9,10], but this is still inadequate for protein assemblies. Here, we present an approach capable of determining structure from diffraction snapshots of symmetric objects to 1/100th of the object diameter, and demonstrate three-dimensional structure recovery to atomic resolution from simulated diffraction snapshots of the satellite tobacco necrosis virus (Fig. 1). This approach can be applied to symmetric objects of any kind, opening the way to the high-resolution study of a wide variety of crystalline and non-crystalline biological and non-biological entities without radiation damage.

The Shannon-Nyquist sampling theorem links the resolution $r$ with which an object can be reconstructed to the number of available snapshots $N_{snaps}$, the object diameter $D$, and the number of elements $N_G$ in the point group of a symmetric object[8]:

$$r = \left( \frac{8\pi^2}{N_G N_{snap}} \right)^{1/3} D \quad . \tag{1}$$

The presence of symmetry can thus substantially increase the achievable resolution. The current experimental concentration on giant viruses[11] (large D), and the scarcity of "useful" single-particle snapshots[12] (small $N_{snap}$) make the exploitation of symmetry crucial for further progress. No reconstruction algorithm capable of operating at the signal-noise ratios expected from single-molecule diffraction has, to date, incorporated object symmetry.

Due to the superior computational efficiency and hence reconstruction capability of non-Bayesian manifold approaches[9], we focus on incorporating this capability into these powerful algorithms[9,10,13]. In brief, these approaches recognize that scattering "maps" a given object orientation to a diffraction snapshot. The collection of all possible orientations in three-dimensional space spans a so-called SO(3) manifold. Scattering maps this manifold to a topologically equivalent compact manifold in the space spanned by the snapshots. We have shown that, to a good approximation, the manifold formed by the snapshots is endowed with the same metric as that of a "symmetric top", loosely speaking a sphere squashed in the direction of the incident beam due to the effect of projection[9]. Such a manifold is naturally described by the Wigner *D*-functions[14], which are intimately related to the elements of the (3x3) rotation matrix[9]. Via so-called Empirical Orthogonal Functions, powerful graph-based algorithms[13,15,16] provide access to the Wigner *D*-functions describing manifolds produced by scattering[9]. The snapshot orientations can then be readily derived[9,10]. We have previously demonstrated this capability at the extremely low signal-to-noise ratios expected for single biological molecules in XFEL diffraction experiments[10], and succeeded in recovering single-molecule structure from experimental cryo-EM snapshots at signal-to-noise ratios an order of magnitude below those needed by standard algorithms[10]. It is the object of this paper to incorporate object symmetry into this approach, and thus enable high-resolution structure recovery from XFEL diffraction snapshots. Here, we show that the theoretically sound and algorithmically powerful Diffusion Map algorithm[15] can be used to recover structure from random snapshots of a symmetric object to high resolution. For concreteness, the discussion is restricted to icosahedral objects, but the approach can be applied to any crystalline or non-crystalline object with symmetry.

The paper is organized as follows. The Results section constructs the eigenfunctions needed to describe manifolds produced by scattering from symmetric objects, describes how symmetry-related ambiguities in orientation recovery may be resolved, and demonstrates structure recovery from simulated diffraction snapshots of a symmetric object to 1/100$^{th}$ of its diameter. For the satellite tobacco necrosis virus used as example, this corresponds to atomic resolution. The Discussions section places our work in the context of ongoing efforts to determine structure from nanocrystals and single particles. The paper concludes with a brief summary of the implications of our work for structure determination by XFEL techniques.

## Results

### Manifolds Produced by Scattering from Symmetric Objects

Diffusion Map describes a manifold in terms of the eigenfunctions of the Laplace-Beltrami operator with respect to an unknown metric[9,13]. In the absence of object symmetry, manifolds produced by scattering are well approximated by a homogeneous metric, with the eigenfunctions of the Laplace-Beltrami operator corresponding to the Wigner *D*-functions[9,10]. In the presence of object symmetry, appropriately symmetrized eigenfunctions are needed. As shown in the Supplementary Information (SI), these can be obtained by summing over the Wigner *D*-functions after operation by the elements of the object point-group, viz.

$$\tilde{D}^j_m(\alpha) = \frac{1}{N_G} \sum_{R_i \in G} O_{R_i} D^j_{m',m}(\alpha) \quad , \tag{2}$$

where $\alpha$ denotes the three numbers collectively representing any rotation, $N_G$ the number of operations $O_{R_i}$ in the point-group *G*, and $D^j_{m',m}(\alpha)$ the (real) Wigner *D*-functions. For the icosahedral group the lowest allowed set of eigenfunctions consists of 13 orthogonal icosahedral functions $\tilde{D}^6_m(\alpha)$, $(-6 \le m \le 6)$. (See Online Methods.) These comprise one non-degenerate (*m* = 0) and six degenerate pairs of eigenfunctions, with the *m* in each pair differing only in sign (Fig. 2).

### Extracting Orientations

Diffusion Map provides the coordinates of each snapshot with respect to Empirical Orthogonal Functions denoted here by $\psi_i$. In principle, each of the first thirteen $\psi_i$ can be identified with a symmetrized eigenfunction $\tilde{D}^6_m(\alpha)$, and the orientation of each snapshot extracted from its coordinates in the space spanned by the thirteen $\psi_i$ . In the absence of substantial noise, inspection of the Diffusion Map eigenvalue spectrum suffices to associate most of the $\psi_i$ with their $\tilde{D}^6_m(\alpha)$ counterparts. More reliably, plots of all snapshot coordinates for different pairs of $\psi_i$ display characteristic patterns, from which each of the 13 $\psi_i$ can be associated with one of the symmetrized Wigner *D*-functions $\tilde{D}^6_m(\alpha)$ (Fig. 3). (The plots corresponding to $m = \pm 3$ and $\pm 5$ are similar. However, snapshots with 5-fold symmetry occur at the center of the $m = \pm 3$ plot and along a circle in $m = \pm 5$ , allowing unambiguous distinction.) In principle, this clears the way to extracting the orientation of each snapshot from its position in the 13-dimensional space.

This is complicated, however, by the presence of symmetry, which introduces degeneracies in the symmetrized eigenfunctions, as outlined above. Clearly, all linear combinations of a normalized degenerate pair are equally acceptable. More precisely, any orthogonal operation on a normalized degenerate pair of eigenfunctions leads to another equivalent pair. Each degenerate pair $\left(\psi_i, \psi_j\right)$ is thus related to its counterpart $\left(\tilde{D}^6_m(\alpha),\, \tilde{D}^6_{-m}(\alpha)\right)$ via an unknown mixing angle $\theta_m$ , and a scaling factor, viz.

$$\begin{pmatrix} \tilde{D}^6_m \\ \tilde{D}^6_{-m} \end{pmatrix} = \frac{1}{\sqrt{13}} \begin{pmatrix} \cos\theta_m & (-1)^{m+1}\sin\theta_m \\ (-1)^m \sin\theta_m & \cos\theta_m \end{pmatrix} \begin{pmatrix} \psi_i \\ \psi_j \end{pmatrix} \tag{3}$$

with the first and second sign options in the mixing matrix corresponding to odd and even values of *m*, respectively (Fig. 4). Additionally, it must be established whether an inversion operation should be inserted on the right side of Eq.

A combination of each of the six degenerate pairs can be thought of as the position of the hour hand on a clock. Ideally, one would like all the clocks to display Greenwich Mean Time (have the same mixing angle $\theta_m$ ). However, the arbitrary orthogonal operations allowed by the presence of degeneracy mean that each clock could show a different "local" time. As orthogonal operations also include inversion, the sense of rotation of each clock could also be reversed. One must therefore determine the local time and the sense of clock rotation in order to relate the Diffusion Map eigenfunctions $\psi_i$ to the symmetrized eigenfunctions $\tilde{D}_m^6(\alpha)$. As described in more detail in Online Methods, this can be accomplished as follows.

First, we show how the linear combination of degenerate pairs - the "local time" or mixing angle $\theta_m$ - can be determined. Consider any snapshot and its conjugate, which is simply its mirror image about an axis perpendicular to the incident beam axis passing through the center of the snapshot. (The choice of line in the plane of the snapshot is equivalent to different orientations of the three-dimensional diffraction volume about the beam axis, which is arbitrary.) A little reflection makes it clear that a snapshot and its conjugate are symmetrically located with respect to the zero of the clock setting (Fig. 4). In other words, the zero of local time is the perpendicular bisector of the line connecting an image to its conjugate. Since the conjugate of an image is easily obtained by a mirror operation through a line in the plane of the snapshots passing through its center, the mixing angle can be readily determined by adding the mirror images of a subset of the snapshots to the dataset before Diffusion Map embedding. Each mixing angle can then be determined to within $\pi$ by the position of the perpendicular bisector of the lines connecting conjugate snapshots (Fig. 4). The remaining $\pi$ ambiguity stems from the sense chosen for the perpendicular bisector, and is resolved later (see below).

Next, we describe how the presence of possible inversions (reversal of the sense of clock rotation) can be determined. Consider a subset of snapshots, and rotate each by a small amount about a central axis perpendicular to its plane to form a new subset of snapshots. Embed the augmented dataset. The sense of rotation can now be determined by observing whether a rotated snapshot leads or trails its unrotated counterpart.

A link is now established between the Diffusion Map eigenfunctions $\psi_i$ and the symmetrized Wigner *D*-functions $\tilde{D}_m^6(\alpha)$, to within a $\pi$ ambiguity in each of the six mixing angles. The snapshot orientations can now be extracted from the $\psi_i$ by a least-squares fit in a straightforward manner, as described in detail in Online Methods. The $\pi$ ambiguity is resolved by performing fits for each of the 64 ($2^6$) possibilities, and selecting the outcome with the lowest residual.

We now demonstrate our approach by reconstructing the structure of an icosahedral virus to 1/100$^{th}$ of its diameter. For the satellite tobacco necrosis virus (STNV, PDB designation: 2BUK) used here, this allows reconstruction to atomic resolution (0.2 nm). The ability of our approach to operate successfully at signal-to-noise ratios as low as $\sim 10^{-2}$ has already been demonstrated with simulated and experimental datasets for a variety of systems[10]. This is ample for dealing with a wide range of weakly scattering biological objects[6,10,17]. The input dataset for the present demonstration, therefore, consists of 1.32 million noise-free simulated diffraction snapshots at an incident photon energy of 12.4 keV to 0.2 nm resolution. (For details, see Online Methods.)

The snapshot orientations were recovered as described above, and used to compile a three-dimensional diffraction volume. Fig. 5 shows a comparison between the exact diffraction volume and that obtained with orientations recovered with our approach. The *R*-factor between the known and reconstructed diffraction volumes is less than 0.14, with approximately half of this stemming from the interpolation needed to place the volume on a regular Cartesian grid (see Online Methods). The electron density was obtained by iterative phasing[18] to recover the atomic-level structure shown in Fig. 1, as described in detail in Online Methods. This completes the demonstration of structure recovery to 1/100$^{th}$ of the object diameter.

## Discussion

Incorporation of object symmetry has proved a powerful tool for recovering structure by single-particle techniques. By enhancing the effective number of snapshots and improving resolution, this promises to play a similarly important role in 3D structure recovery by XFEL methods. In terms of resolution expressed as a fraction of the object diameter, our approach is comparable with the best achieved by cryo-EM approaches[19], but without requiring phase information. Combined with its superior noise robustness[10,12], our approach offers a vital route to determining high-resolution structure at signal levels expected from single macromolecules in XFEL experiments[6,10,17]. For biological entities in particular, this is essential for obtaining "biologically relevant" information.

XFEL experiments to obtain snapshots from individual biological objects are in progress[11,20]. Analysis of the only publicly available XFEL dataset on viruses[20], however, reveals the overwhelming influence of experimental artifacts, such as variations in beam intensity, position and inclination, and limitations due to detector dynamic range and nonlinearities. These have so far prevented three-dimensional structure recovery. Judging by the rapid progress in XFEL-based nanocrystallography[2], improved datasets are expected to emerge quickly. Our approach thus represents a vital and timely tool for high-resolution structure recovery from symmetric biological and non-biological single particles before the onset of radiation damage.

More generally, our approach can be applied also to structure recovery by XFEL-based nanocrystallographic methods. Traditional indexing approaches, combined with Monte Carlo integration techniques have provided impressive first results[1-3]. However, issues such as the so-called "twinning ambiguity" and the effect of variations in nanocrystal size and shape have so far eluded resolution. The incorporation of symmetry into manifold-based approaches offers a potentially powerful route to resolving these important problems. If successful, this would substantially extend the reach of XFEL-based nanocrystallography.

**Conclusions**

We have presented the first approach capable of extracting high-resolution structure from diffraction snapshots of symmetric objects, and demonstrated structure recovery to 1/100$^{th}$ of the object diameter. Together with its previously demonstrated capability to operate at exceptionally low signal-to-noise ratios[10], this opens the way to the study of individual biological entities before the onset of significant radiation damage. Our approach also offers the possibility to apply powerful graph-theoretic techniques to the study of crystalline objects, with the potential to extract more information from the rich and rapidly growing body of nanocrystallographic data.

**Acknowledgments**

We are grateful to Dimitrios Giannakis and Dilano Saldin for valuable discussions. This research was supported by: the US Dept. of Energy, Office of Science, Basic Energy Sciences under award DE-FG02-09ER16114 (AO: overall design); the US National Science Foundation under awards MCB-1240590 (AH: algorithm development), CCF-1013278 and CNS-0968519 (AD and RMD: GPU algorithms); and the UWM Research Growth Initiative (RF: algorithm design).

**Figure Captions**

**Fig. 1**: **Recovered structure of the satellite tobacco necrosis virus to atomic resolution.**

**(a)** The three-dimensional electron density extracted from 1.34 million diffraction snapshots of unknown orientation, demonstrating structure recovery to a resolution 1/100$^{th}$ of the object diameter (0.2nm).

**(b)** A slice of the electron density showing atomic resolution. The known structure is shown as a ball-and-sticks model with no adjustments.

**Fig. 2**: **Eigenvalue spectra of the Laplace-Beltrami operator.**

**(a)** Spectrum for icosahedral Wigner *D*-functions.

**(b)** Spectrum obtained from the manifold produced by diffraction snapshots of the satellite tobacco necrosis virus.

Note the close agreement between the two spectra.

**Fig. 3**: **Scatter plots to identify eigenfunctions.**

These, together with the distribution of 5-fold symmetric snapshots allow unambiguous association of the Diffusion Map eigenvectors $\psi_i$ with their counterparts among the icosahedral Wigner *D*-functions $\tilde{D}_m^6$.

**(a)** $\psi_i$ vs. $\psi_j$ plots of snapshot coordinates obtained from Diffusion Map.

**(b)** $\tilde{D}_m^6$ vs. $\tilde{D}_{-m}^6$ plots for randomly sampled points in the space of orientations.

**Fig. 4**: **Schematic diagram showing the effect of the mixing angle $\theta_m$ between a pair of normalized degenerate eigenfunctions.**

The zero of the mixing angle is the given by the perpendicular bisector of the line connecting a snapshot to its conjugate. The sense of clock rotation can be determined from the position of a snapshot rotated by a few degrees about the beam axis.

**Fig. 5**: **Comparison of exact and recovered diffraction volumes.**

**(a)** A slice through the exact three-dimensional diffraction volume.

**(b)** Same slice through the diffraction volume obtained with orientations deduced by the approach described in this paper.

## References

1. Chapman, H.N. et al. Femtosecond X-ray protein nanocrystallography. *Nature* **470**, 73-77 (2011).
2. Boutet, S. et al. High-resolution protein structure determination by serial femtosecond crystallography. *Science* **337**, 362-364 (2012).
3. Redecke, L. et al. Natively inhibited Trypanosoma brucei cathepsin B structure determined by using an X-ray laser. *Science* **339**, 227-230 (2013).
4. Neutze, R., Wouts, R., van der Spoel, D., Weckert, E. & Hajdu, J. Potential for biomolecular imaging with femtosecond X-ray pulses. *Nature* **406**, 752-757 (2000).
5. Schwander, P., Fung, R., Phillips, G.N. & Ourmazd, A. Mapping the conformations of biological assemblies. *New Journal of Physics* **12**, 1-15 (2010).
6. Fung, R., Shneerson, V., Saldin, D.K. & Ourmazd, A. Structure from fleeting illumination of faint spinning objects in flight. *Nature Physics* **5**, 64-67 (2009).
7. Loh, N.T. & Elser, V. Reconstruction algorithm for single-particle diffraction imaging experiments. *Phys Rev E.* **80**, 026705 (2009).
8. Moths, B. & Ourmazd, A. Bayesian algorithms for recovering structure from single-particle diffraction snapshots of unknown orientation: a comparison. *Acta Cryst A* **A67**, 481-486 (2011).
9. Giannakis, D., Schwander, P. & Ourmazd, A. The symmetries of image formation by scattering. I. Theoretical framework. *Optics Express* **20**, 12799 - 12826 (2012).
10. Schwander, P., Yoon, C.H., Ourmazd, A. & Giannakis, D. The symmetries of image formation by scattering. II. Applications. *Optics Express* **20**, 12827-12849 (2012).
11. Seibert, M.M. et al. Single mimivirus particles intercepted and imaged with an X-ray laser. *Nature* **470**, 78-81 (2011).
12. Yoon, C.H. et al. Unsupervised classification of single-particle X-ray diffraction snapshots by spectral clustering. *Optics Express* **19**, 16542-16549 (2011).
13. Coifman, R.R. et al. Geometric diffusions as a tool for harmonic analysis and structure definition of data: diffusion maps. *Proc Natl Acad Sci U S A* **102**, 7426-7431 (2005).
14. Hu, B.L. Scalar waves in the mixmaster universe. I. The Helmholtz Equation in a fixed background *Phys. Rev. D* **8**, 1048-1060 (1973).
15. Coifman, R. & Lafon, S. Diffusion Maps. *Appl. Comput. Harmon. Anal.* **21**, 5-30 (2006).
16. Belkin, M. & Niyogi, P. Laplacian eigenmaps for dimensionality reduction and data representation. *Neural Computation* **15**, 1373-1396 (2003).
17. Shneerson, V.L., Ourmazd, A. & Saldin, D.K. Crystallography without crystals. I. The common-line method for assembling a three-dimensional diffraction volume from single-particle scattering. *Acta Crystallogr A* **64**, 303-315 (2008).
18. Oszlanyi, G. & Suto, A. *Ab initio* structure solution by charge flipping. *Acta Cryst. A* **60**, 134-141 (2004).
19. Zhang, X. et al. Three-dimensional structure and function of the Paramecium bursaria chlorella virus capsid. *Proc Natl Acad Sci U S A* **108**, 14837-14842 (2011).
20. Kassemeyer, S. et al. Femtosecond free-electron laser x-ray diffraction data sets for algorithm development. *Optics Express* **20**, 4149-4158 (2012).

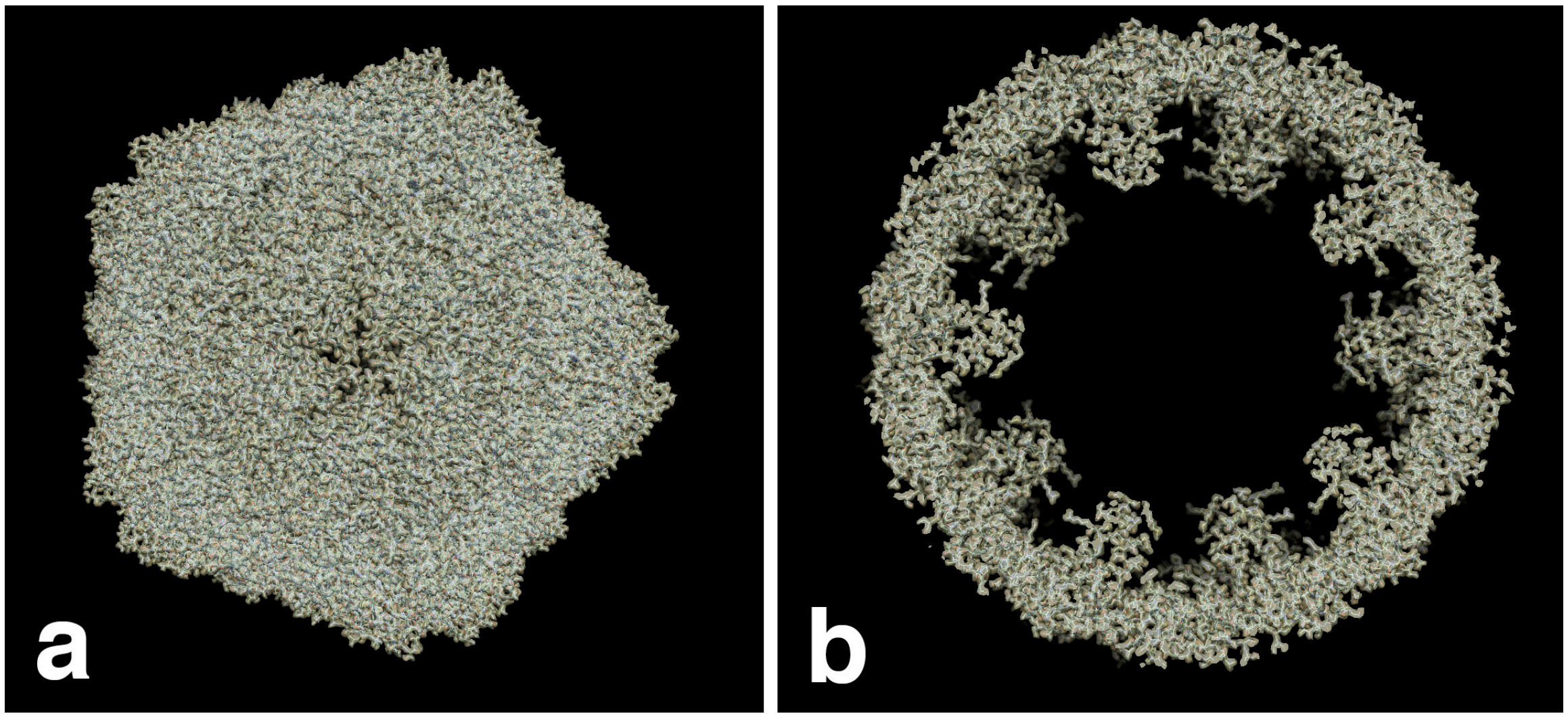
a
b

**Fig. 1**

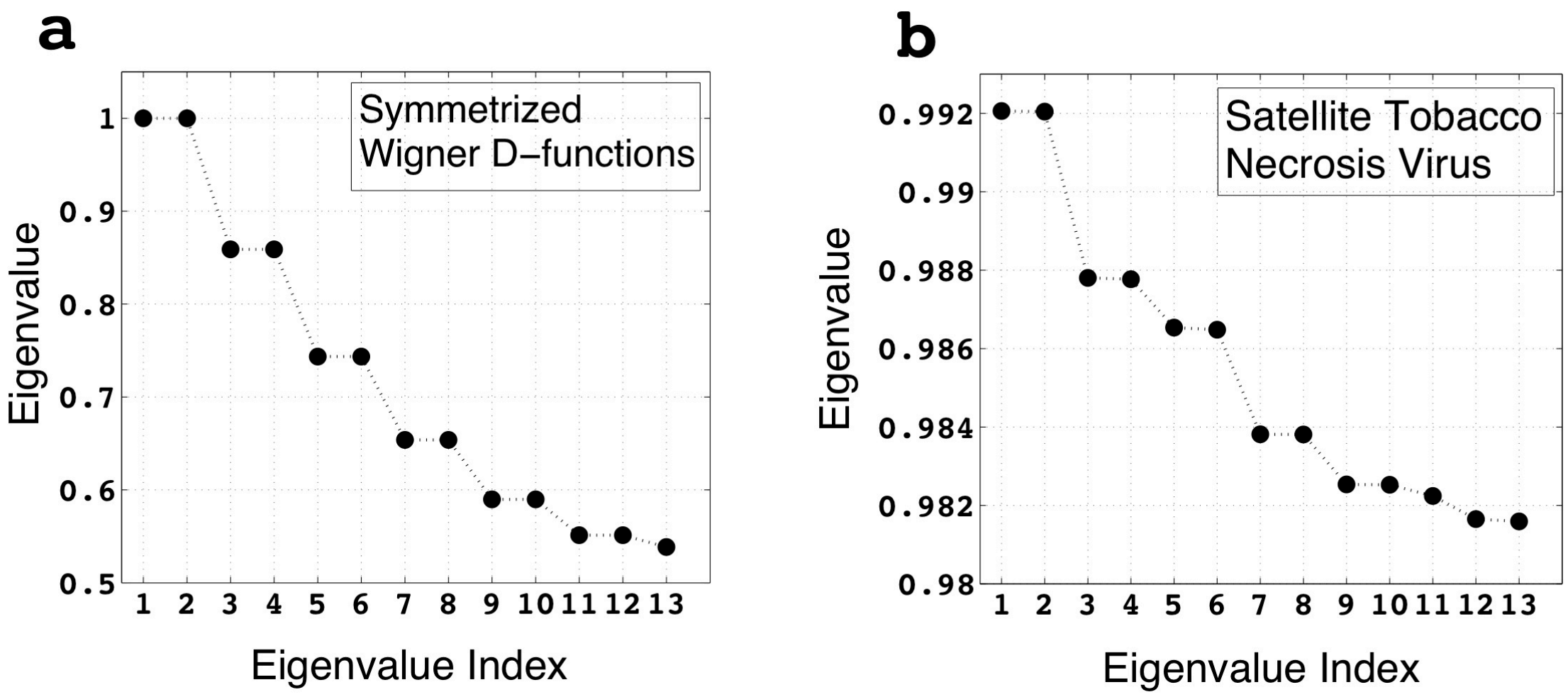
a
Symmetrized
Wigner D−functions
Eigenvalue
1
0.9
0.8
0.7
0.6
0.5
1 2 3 4 5 6 7 8 9 10 11 12 13
Eigenvalue Index
b
Satellite Tobacco
Necrosis Virus
Eigenvalue
0.992
0.99
0.988
0.986
0.984
0.982
0.98
1 2 3 4 5 6 7 8 9 10 11 12 13
Eigenvalue Index

**Fig. 2**

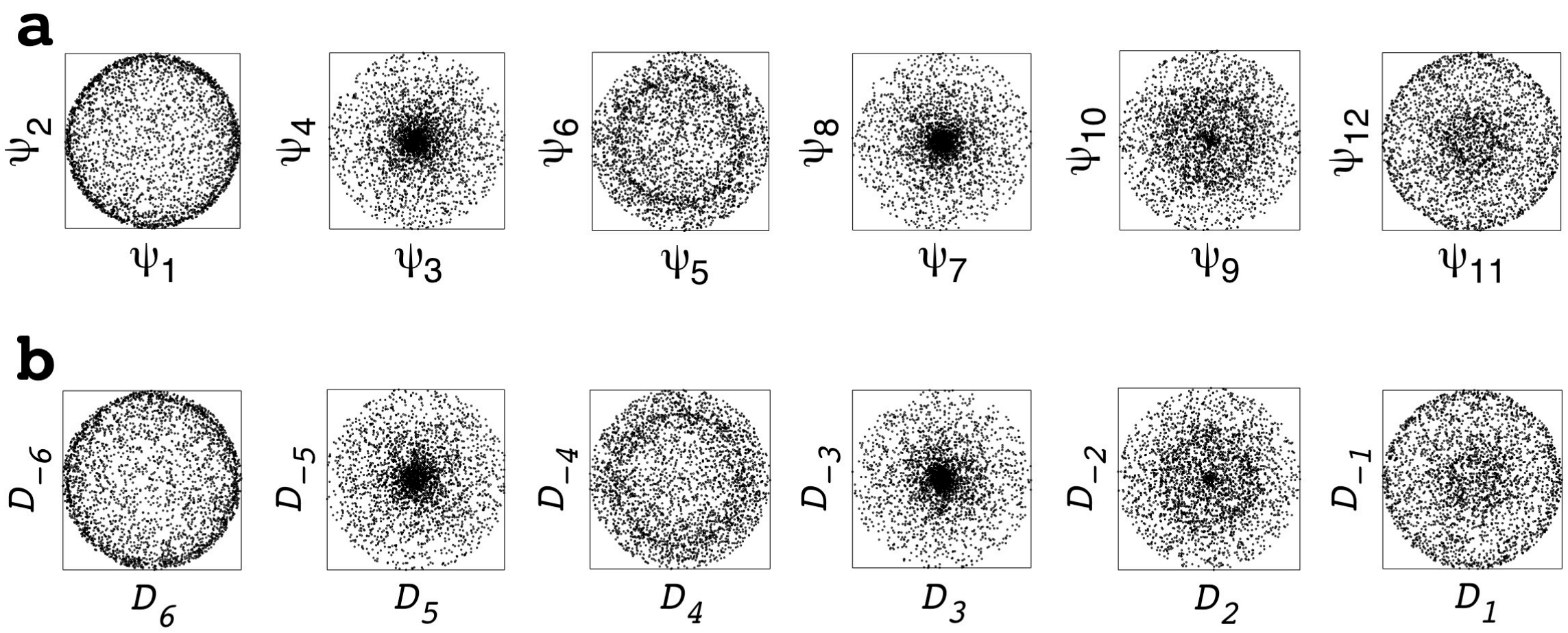

**Fig. 3**

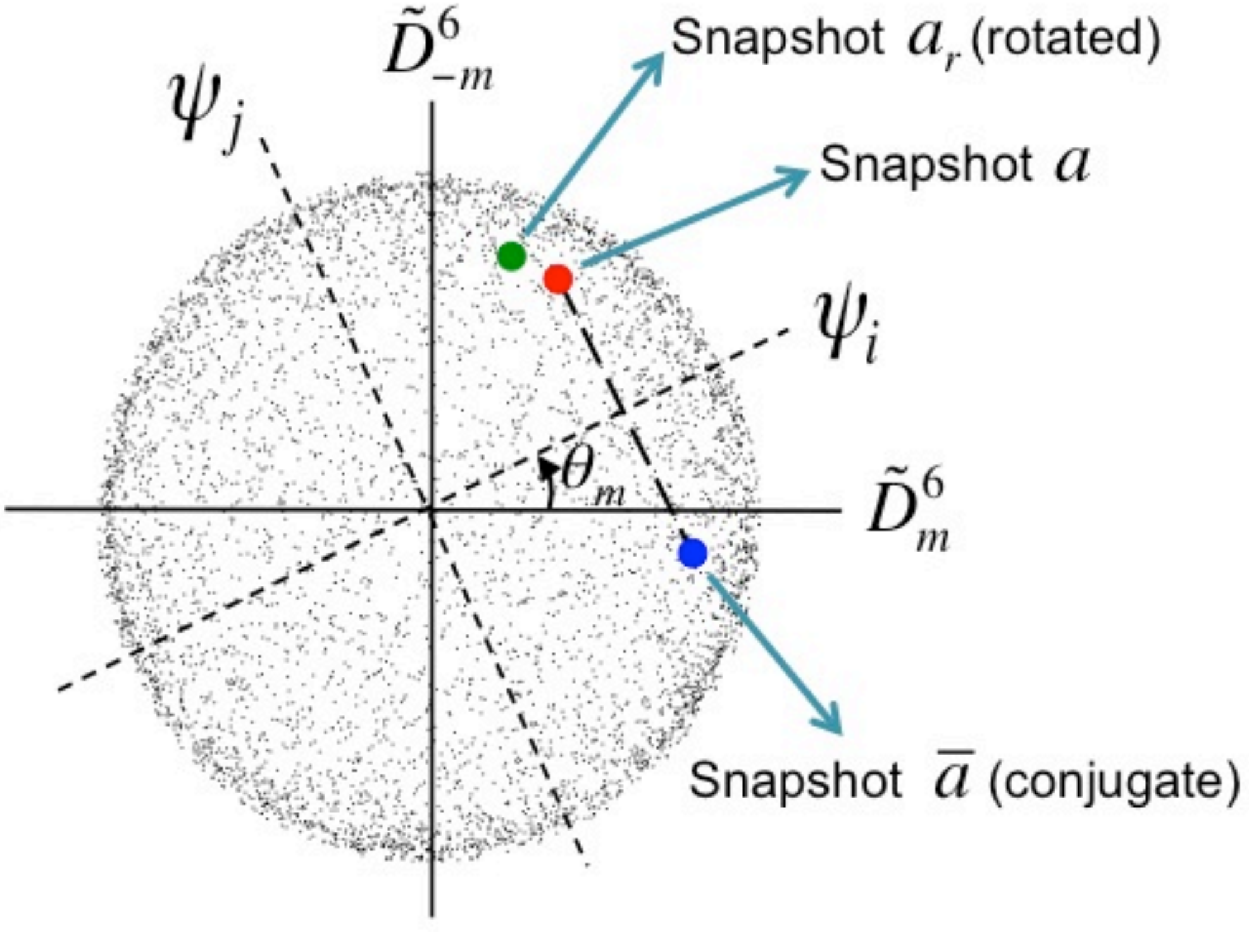

**Fig. 4**

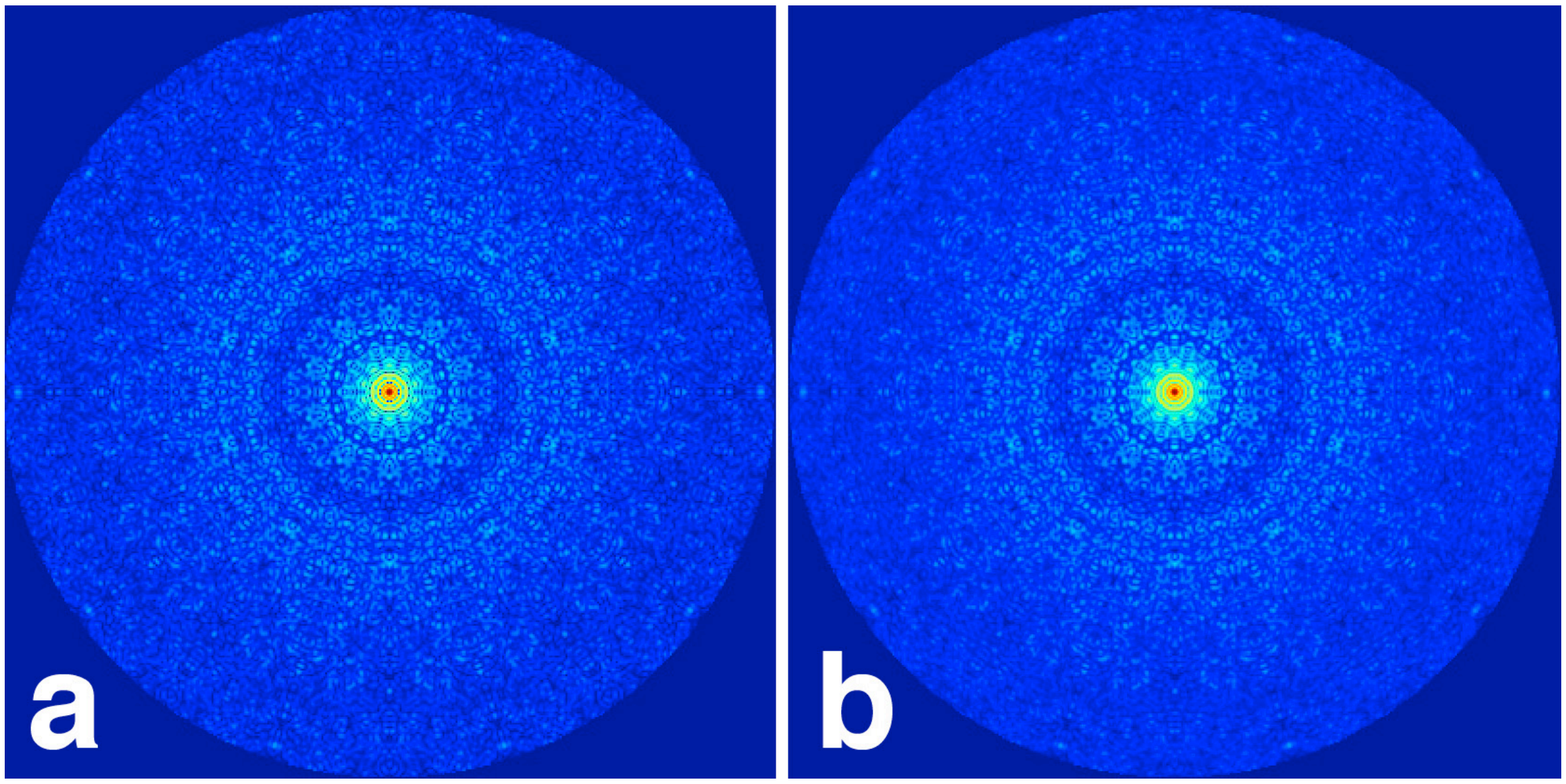

**Fig. 5**